# Six Strange Facts about our First Interstellar Guest, `Oumuamua

Abraham Loeb

`Oumuamua is nothing like we expected or seen before.

———

By Abraham Loeb on November 18, 2018

On October 19, 2017, the first interstellar object, `Oumuamua, was [discovered](#)[1] by the [Pan-STARRS survey](#)[2] within a sixth of the Earth-Sun distance. The experience was similar to having a surprise guest for dinner from another country. By examining this guest we can learn about the culture of that country without the need to travel there. Given the vast distance involved, it would have taken us a hundred thousand years to visit the nearest star using conventional chemical rockets.

Surprisingly, our first interstellar guest appeared to be weird and unlike anything we have seen before. By the time we realized it, the guest was already out the door with its image fading into the dark street, so we did not have a chance to get a second look at its mysterious qualities.  Below is a list of six peculiarities exhibited by `Oumuamua:

1. Assuming that other planetary systems resemble the Solar System, Pan-STARRS should not have discovered any interstellar rock in the first place. In a [paper published a decade ago](#)[3], we predicted an abundance of interstellar asteroids that is smaller by many (2-8) orders of magnitude than needed to explain the discovery of `Oumuamua as a member of a random population of objects. In other words, the population of interstellar objects is far greater than expected. Each star in the Milky-Way [needs to eject](#)[4] about $10^{15}$ such objects during its lifetime to account for the inferred population, much more than anticipated based on the Solar System. Thus, the nurseries of `Oumuamua-like objects must be different from the familiar ones.
2. `Oumuamua [originated from a very special frame of reference](#)[5], near the so-called Local Standard of Rest (LSR), which is defined by averaging the random motions of all the stars in the vicinity of the Sun. Only one star in five hundred is as slow as `Oumuamua in that frame. The LSR is the ideal frame for camouflage, namely for hiding the origins of an object and avoiding its association with any particular star - since stars typically move in that frame. The relative motion between `Oumuamua and the Sun reflects the motion of the Sun relative to the LSR. `Oumuamua is like a buoy sitting at rest on the surface of the ocean, with the Solar System running into it like a fast ship. Could there be an array of buoys that serves as a network of relay stations or road posts, defining the average Galactic frame of reference in interstellar space?
3. Most interstellar asteroids are expected to be ripped apart from their parent star in the outskirts of their birth planetary system (such as the Oort cloud in the Solar System which extends to a hundred thousand times the Earth-Sun separation), where they are most loosely bound to the star's gravity. At these outskirts, they can be removed with a small velocity nudge of less than a kilometer per second, in which case they will inherit the speed of their host star relative to the LSR. If `Oumuamua came from a typical star, it must have been ejected with an unusually large velocity kick. To make things more unusual, its kick should have been equal and opposite to the velocity of its parent star relative to the LSR, which is about twenty kilometers per second for a typical star like the Sun. The dynamical origin of `Oumuamua is extremely rare no matter how you look at it. This is surprising, since the first foreign guest to a dinner party should be statistically common (especially given the larger than usual population inferred in the first point above).

4. We do not have a photo of `Oumuamua but its brightness owing to reflected sunlight varied by a factor of 10 as it rotated periodically every eight hours. This implies[6] that `Oumuamua has an extreme shape with its length at least 5-10 times larger than its projected width. Moreover, an analysis[7] of its tumbling motion concluded that it would be at its highest excitation state as expected from its tumultuous journey, if it has a pancake-like geometry. The inferred shape is more extreme than for all asteroid previously seen in the Solar System, which have an axes ratio of at most 3.
5. The Spitzer Space Telescope did not detect[8] any heat in the form of infrared radiation from `Oumuamua. Given the surface temperature dictated by `Oumuamua's trajectory near the Sun, this sets an upper limit of hundreds of meters on its size. Based on this size limit, `Oumuamua must be unusually shiny with a reflectance that is at least ten times higher than exhibited by Solar System asteroids.
6. The trajectory of `Oumuamua deviated[9] from that expected based on the Sun's gravity alone. The deviation is small (a tenth of a percent) but highly statistically significant. Comets exhibit such a behavior when ices on their surface heat up from solar illumination and evaporate, generating thrust through the rocket effect. The extra push for `Oumuamua could have originated by cometary outgassing if at least a tenth of its mass evaporated. This massive evaporation would have naturally led to the appearance of a cometary tail, but none was seen. The Spitzer telescope observations place tight limits on any carbon-based molecules or dust around `Oumuamua, and rule out the possibility that normal cometary outgassing is at play (unless it is composed of pure water). Moreover, cometary outgassing would have changed the rotation period of `Oumuamua[10] and no such change was observed. Altogether, `Oumuamua does not appear to be a typical comet nor a typical asteroid, even though it represents a population that is far more abundant than expected.

The extra push exhibited by `Oumuamua's orbit could not have originated from a breakup into pieces because such an event would have provided an impulsive kick unlike the continuous push that was observed. If cometary outgassing is ruled out and the inferred excess force is real, only one possibility remains - an extra push due to radiation pressure from the Sun[11]. In order for this push to be effective, `Oumuamua needs to be less than a millimeter thick but with a size of at least twenty meters (for a perfect reflector), resembling a lightsail of artificial origin. In this case `Oumuamua would resemble the solar sail demonstrated by the Japanese mission IKAROS[12] or the lightsail contemplated for the Starshot initiative[13]. An artificial origin offers the startling possibility that we discovered "a message in a bottle", following years of failed searches for radio signals from alien civilizations. Reassuringly, such a lightsail would survive collisions with interstellar atoms and dust as it travels throughout the Galaxy.

In contemplating the possibility of an artificial origin, we should keep in mind what Sherlock Holmes said: "when you have excluded the impossible, whatever remains, however improbable, must be the truth". The Kepler satellite [revealed](#)[14] that about a quarter of all the stars in the Milky Way have a habitable planet of the size of the Earth, with the potential to have liquid water on its surface and the chemistry of life as we know it. It is therefore conceivable that interstellar space is full of artificially-made debris, either in the form of devices that serve a purpose on a reconnaissance mission or defunct equipment. However, to validate the exotic artificial origin for `Oumuamua we need more data. As Carl Sagan said "extraordinary claims require extraordinary evidence".

Interestingly, the possibility of a targeted mission adds some explanatory power. It is unlikely that $10^{15}$ solar sails are launched per star to make up a random population of `Oumuamua-like objects. This would require the unreasonable rate of a launch every five minutes from a planetary system even if all civilizations live as long as the full lifetime of the Milky Way galaxy. Instead, the required numbers could be reduced dramatically if `Oumuamua-like objects do not sample all possible orbits randomly but rather follow special orbits that dive into the innermost, habitable regions of planetary systems like the Solar System.

`Oumuamua moves too fast for our chemical rockets to catch up with it now without [gravitational assist from planets](#)[15]. But since it would take `Oumuamua thousands of years to leave the solar system, getting a closer look of it through a flyby remains a possibility even if we were to develop new technologies for faster space travel within a decade or two. Interestingly, some interstellar objects which pass close to Jupiter can lose energy and [get captured by the Solar System](#)[16]. These are dinner guests who bumped into a wall on their way out and stayed around after dinner. The Sun-Jupiter system acts as a fishing net. If we can identify trapped interstellar objects through their unusual bound orbits with unusually high inclinations relative to the Solar System plane, we could design missions to visit them and learn more about their nature.

Alternatively, we can wait for the next interstellar guest to show up. Within a few years, the [Large Synoptic Survey Telescope (LSST)](#)[17] will become operational and be far more sensitive to the detection of `Oumuamua-like objects. It should therefore discover many `Oumuamua-like objects within its first year of operation. If it does not find any, we will know that `Oumuamua was special and that we must chase this guest down the street in order to figure out its origin.

Studying interstellar objects resembles my favorite activity when walking along the beach with my daughters. We enjoy picking up sea shells that were swept ashore and learning about their different origins. Every now and then, we find a plastic bottle that indicates an artificial origin. Similarly, astronomers should examine any object that enters the Solar System and study its properties. There is no doubt that the six peculiar features of `Oumuamua usher in a new era of space archaeology.

## ABOUT THE AUTHOR

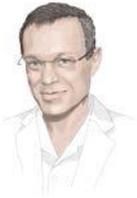

**Abraham Loeb**

Abraham (Avi) Loeb is chair of the astronomy department at Harvard University, founding director of Harvard's Black Hole Initiative and director of the Institute for Theory and Computation at the Harvard-Smithsonian Center for Astrophysics. He chairs the Board on Physics and Astronomy of the National Academies and the advisory board for the Breakthrough Starshot project.

Credit: Nick Higgins